\documentclass[12pt,a4paper,final]{iopart}

\usepackage{iopams}  
\usepackage{graphicx}

\usepackage{amssymb}
\usepackage{amscd}	
\usepackage{enumerate}
\usepackage{subfigure}
\usepackage{xcolor}
\usepackage{color}
\usepackage[margin=1cm]{caption}

\newcommand{\eqref}[1]{(\ref{#1})}

\eqnobysec

\begin{document}





\title{Memory-assisted measurement-device-independent quantum key distribution}

\author{Christiana Panayi$^{1}$, Mohsen Razavi $^{1}$, Xiongfeng Ma$^{2}$, Norbert L\"utkenhaus$^{3}$}
\address{$^1$School of Electronic and Electrical Engineering, University of Leeds, Leeds, UK}
\address{$^2$Center for Quantum Information, Institute for Interdisciplinary Information Sciences, Tsinghua University, Beijing, China}
\address{$^3$Institute for Quantum Computing, University of Waterloo, Waterloo, Canada}
\ead{py10cp@leeds.ac.uk, m.razavi@leeds.ac.uk, xma@tsinghua.edu.cn, nlutkenhaus@uwaterloo.ca}


\begin{abstract}
A protocol with the potential of beating the existing distance records for conventional quantum key distribution (QKD) systems is proposed. It borrows ideas from quantum repeaters by using memories in the middle of the link, and that of measurement-device-independent QKD, which only requires optical source equipment at the user's end. For certain memories {with short access times,} our scheme allows a higher repetition rate than that of quantum repeaters {with single-mode memories}, thereby requiring lower coherence times. By accounting for various sources of nonideality, such as memory decoherence, dark counts, misalignment errors, and background noise, as well as timing issues with memories, we develop a mathematical framework within which we can compare QKD systems with and without memories. In particular, we show that with the state-of-the-art technology for quantum memories, it is possible to devise memory-assisted QKD systems that, at certain distances of practical interest, outperform current QKD implementations.    
\end{abstract}

\vspace{2pc}
\maketitle

\section{Introduction}

Despite all commercial \cite{Commercial_QKD} and experimental achievements in QKD \cite{Wang:260kmQKD:2012,Sasaki:TokyoQKD:2011, secoqc, Townsend_QI_home_2011,Shields.PRX.coexist, Tittel:expMDIQKD_PRL2013, Silva_expMDIQKD_PRA2013, Pan_expMDIQKD_PRL2013, HKLo:expMDIQKD_2013}, reaching arbitrarily long distances is still a remote objective. The fundamental solution to this problem, i.e., quantum repeaters, is known for over a decade. {From early proposals by Briegel {\em et al.} \cite{Zoller_Qrepeater_98} to {the latest no-memory versions} \cite{Munro:NatPhot:2012, Azuma:All_optical_QR_2013, Liang:NoMemRep_2013}, quantum repeaters, typically, rely on highly efficient quantum gates comparable to what we may need for future quantum computers. While the progress on that ground may take some time before such systems become functional, another approach based on {\em probabilistic} gate operations was proposed by Duan and co-workers \cite{DLCZ_01}, which could offer a simpler way of implementing quantum repeaters for moderate distances of up to around 1000~km. The latter systems require quantum memory modules with high coupling efficiencies to light {\em and} with coherence times exceeding the transmission delays, which are yet to be achieved together.} In this paper, we propose a protocol that, {although is not as scalable as quantum repeaters,} for certain classes of memories, relaxes, to some extent, the harsh requirements on memories' coherence times, thereby paving the way for the existing technologies to beat the highest distance records achieved for {no-memory} QKD links \cite{Wang:260kmQKD:2012}. The idea behind our protocol was presented in \cite{Panayi_ICQNM12}, and {independent work has also been reported in} \cite{Brus:MDIQKD-QM_2013}. This work proposes additional practical schemes and rigorously analyses them under realistic conditions.

Our protocol relies on concepts from quantum repeaters, on the one hand, and the recently proposed measurement-device-independent QKD (MDI-QKD), on the other. The original MDI-QKD \cite{Lo:MIQKD:2012} relies on sending encoded photons by the users to a middle site at which a Bell-state measurement (BSM) is performed. {One major practical advantage of MDI-QKD is that this BSM can be done by an {\em untrusted} party, e.g., the service provider, which makes MDI-QKD resilient to detector attacks, e.g., time-shift, remapping, and blinding attacks  \cite{Qi:TimeShift:2007, Zhao:TimeshiftExp:2008, Fung:Remap:07,HKLO_PhaseRemap_NJP2010, Lydersen:Hacking:2010, Wiechers:AftergateAttack:2011, Weier:DeadtimeAttack:2011, Jain:AttackExp:2011}. The security is then guaranteed by the reverse EPR protocol \cite{Biham:ReverseEPR:1996}. Another practical advantage is that this BSM does not need to be a perfect measurement, but even a partial imperfect BSM implemented by linear optical elements can do the job.} In our scheme, by using two quantum memories at the middle site, we first store the state of the transmitted photons in the memories, and perform the required BSM, only when both memories are loaded. {In that sense, our memory-assisted MDI-QKD is similar to a single-node quantum repeater, except that there is no memories at the user's end.} This way, similar to quantum repeaters, we achieve a rate-versus-distance improvement as compared to the MDI-QKD schemes proposed in \cite{Lo:MIQKD:2012, MXF:MIQKD:2012, MDIQKD_finite_PhysRevA2012, Braunstein:MIQKD:2012}, or other conventional QKD systems that do not use quantum memories. 


There is an important distinction between our protocol and a conventional quantum repeater system {that relies on single-mode memories}. In such a quantum repeater link, which relies on initial entanglement distribution among neighbouring nodes, the repeat period for the protocol is mainly dictated by the transmission delay for the shortest segment of the repeater system \cite{Razavi.Lutkenhaus.09,Razavi_SPIE}. In our scheme, however, the repeat period is constrained by the writing time, including the time needed for the herald/verification process, into memories. This implies that using sufficiently fast memories, {i.e., with short writing times,} one can run our scheme at a faster rate than that of a quantum repeater, thereby achieving higher key generation rates, as compared to conventional QKD links, and at lower coherence times, as compared to probabilistic repeater systems. This increase in clock rate is what our proposal shares with the recently proposed third generation of quantum repeaters, which use quantum error correction codes to compensate for loss and errors, thus also being able to speed up the clock rate to local processing times \cite{Munro:NatPhot:2012}. The need for long coherence times remains one of the key challenges in implementing the first generations of quantum repeaters before the latest no-memory quantum repeater proposals can be implemented.

The above two benefits would offer a midterm solution to the problem of long-distance QKD. While our scheme is not scalable the same way that quantum repeaters are, it possibly allows us to use the existing technology for quantum memories to improve the performance of QKD systems. In the absence of fully operational quantum repeater systems, our setup can fill the gap between theory and practice and will become one of the first applications of realistic quantum memories in quantum communications. 

{It is worth mentioning that the setups we propose here are compatible with different generations of hybrid quantum-classical (HQC) networks \cite{Razavi_IWCIT12}. In such systems, home users are not only able to use broadband data services, but they can also use quantum services such as QKD. MDI-QKD offers a user-friendly approach to the access part of such networks as the end users only require source equipment. Whereas, in the first generation of HQC networks, the service provider may only facilitate routing services for quantum applications, in the future generations, probabilistic, deterministic, and eventually no-memory quantum repeaters constitute the quantum core of the network. In each of these cases, our setups are extensible and compatible with forthcoming technologies for HQC networks.}

The rest of the paper is structured as follows. In Section II, we describe our proposed schemes and the modelling used for each component therein. Section III presents our key rate analysis, followed by some numerical results in Section IV. Section V concludes the paper.


\section{System Description}
\label{Sec:System}

{Our scheme relies on ``loading'' quantum memories (QMs) with certain, unknown, states of light. This loading process needs to be heralding, that is, by the end of it, we should learn about its success. Within our scheme, two types of memories can be employed, which we refer to by {\em directly} versus {\em indirectly} heralding QMs. Some QMs can operate in both ways, while some others are more apt to one than the other. By directly heralding memories we refer to the class of memories to which we can directly transfer the state of a photon {\em and} we can verify---without revealing or demolishing the quantum state---whether this writing process has been successful. An example of such memories is a trapped atom in an optical cavity \cite{MIT-NU}. In the case of indirectly heralding memories, a direct writing-verification scheme may not exist. Instead, we assume that we can entangle a photonic state with the state of such QMs \cite{DLCZ_01, Razavi.DLCZ.06, KBBBCDK_03, Kuzmich_memory_05, Zhao:Robust:2007, Pan:NatPhys:2012, Rempe:Nature:2012}, and later, by doing a measurement on the photon, we can effectively achieve a heralded writing into the memory. These two approaches of writing cover most relevant practical examples to our scheme.}

\begin{figure}[hbt]
\centering
{\includegraphics[width = 12.6cm]{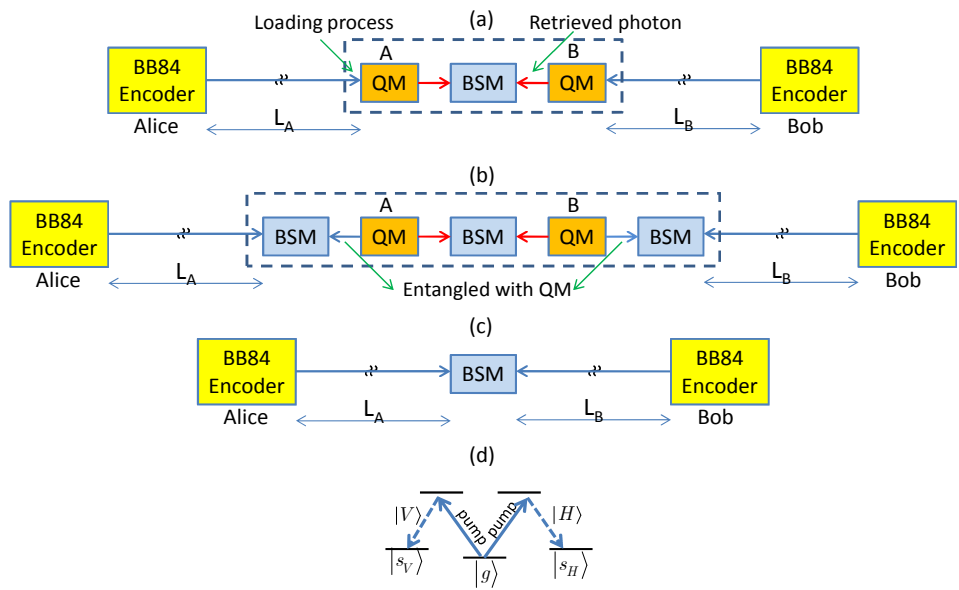}}
\caption{(a) MDI-QKD with directly heralding quantum memories. Alice and Bob use the efficient BB84 protocol to encode and send pulses to their respective QM in the middle of the link. At each round, each QM attempts to store the incoming pulse. Once they are both loaded, we retrieve the QMs' states and perform a BSM on the resulting photons. (b) MDI-QKD with indirectly heralding quantum memories. At each round, an entangling process is applied to each QM, which would generate a photon entangled, in polarization, with the QM. These photons interfere at the BSM modules next to the QMs with incoming pulses from the encoders. As soon as one of these BSMs succeeds, we stop the entangling process on the corresponding QM, and wait until both QMs are ready for the middle BSM operation. In this case, QMs are not required to be heralding; a trigger event is declared by the success of the BSM located between the QM and the respective encoder. (c) The original MDI-QKD protocol \cite{Lo:MIQKD:2012}. {(d) One possible energy-level configuration for a QM suitable for polarization encoding.}}
\label{Fig:figure1_a_b_c}
\end{figure}

The scheme for directly heralding memories works as follows \cite{Panayi_ICQNM12,Brus:MDIQKD-QM_2013}; see figure~\ref{Fig:figure1_a_b_c}(a). The two communicating parties, Alice and Bob, send BB84 encoded pulses \cite{BB_84}, by either single-photon or weak laser sources, towards QM units located in the middle of the link. Each QM stores a photon in a possibly probabilistic, but {\em heralding}, way. Once both memories are loaded, we retrieve their states and perform a BSM on the corresponding photons. A successful BSM indicates some form of correlation between the transmitted bits by Alice and Bob. 

We can easily extend the above idea to the case of indirectly heralding memories. An additional BSM, on each side, along with an entangling process between photons and QMs, can replace the verification process needed for directly heralding memories. In this case, see figure~\ref{Fig:figure1_a_b_c}(b), a successful BSM between the transmitted photon by the users and the one entangled with the QM, would effectively herald a successful loading process, that is, the state of the QM is correlated with the quantum state sent by the users. 

{In order to entangle a QM with a photon, one can think of two standard ways. One approach would be to generate a pair of entangled photons, e.g., by using spontaneous parametric down-converters \cite{Kuklewicz:BrightEPR, MIT_polEntgSource}, and then store one of the photons in the memory and use the other one for interference with the incoming photon sent by the user. While this approach is not fully heralding (because we cannot be sure of the absorption of the locally generated photon by the memory), it is still a viable option for highly efficient writing procedures.
Another approach to entangle a photon with a memory, which this paper is mainly concerned with, is to start from the memory and generate a photon entangled with the memory by driving certain transitions in the memory \cite{Rempe:Nature:2012, DLCZ_01}. With entangling times as short as 300~ps reported in the literature \cite{Walmsley:PRL:2010}, high repetition rates are potentially achievable for indirectly heralding memories.}  

In either approach, it is possible to have multiple-excitation effects, which can cause errors in our setup. In this paper, for readability reasons, we make the simplifying assumption of having only single excitations in the memories, and address the multiple excitation effect in a separate publication \cite{LoPiparo:2014}. Furthermore, here we only consider the polarization entanglement. The extension to other types of entanglement is straightforward and will be dealt with in forthcoming publications.

{Under all above assumptions, suppose once we entangle the memory $A$ with a single photon $P$, the joint state of the two is given by
\begin{equation}
\label{jointent}
\frac{1}{\sqrt{2}}[|s_H\rangle_A |H\rangle_P + |s_V\rangle_A |V\rangle_P],
\end{equation}
where $|H\rangle_P$ and $|V\rangle_P$, respectively, represent horizontally and vertically polarized single photons, and $|s_H\rangle_A$ and $|s_V\rangle_A$ are the corresponding memory states; see figure~\ref{Fig:figure1_a_b_c}(d). {In equation~\eqref{jointent}, the conditional state of the photon, knowing the memory state, has the same form as in BB84.}  Each leg of figure~\ref{Fig:figure1_a_b_c}(b), from the user end to the respective QM, is then similar to an asymmetric setup of the original MDI-QKD scheme as depicted in figure~\ref{Fig:figure1_a_b_c}(c). The working of the system in figure~\ref{Fig:figure1_a_b_c}(b) will then follow that of the original MDI-QKD. We will use this similarity in our analysis of the system in figure~\ref{Fig:figure1_a_b_c}(b).}

The main advantage of our scheme as compared to the original MDI-QKD, in figure~\ref{Fig:figure1_a_b_c}(c), is its higher resilience to channel loss and dark count. In the no-memory MDI-QKD, both pulses, sent by Alice and Bob, should survive the path loss before a BSM can be performed. The key generation rate then scales with the loss in the entire channel. In our scheme, each pulse still needs to survive the path loss over half of the link, but this can happen in different rounds for the signal sent by Alice as compared to that of Bob. We therefore achieve the quantum repeater benefit in that the key generation rate, in the symmetric case, scales with the loss over half of the total distance. Moreover, in the case of directly heralding memories, our scheme is almost immune against dark counts \cite{Brus:MDIQKD-QM_2013}. This is because the measurement efficiency in the BSM module is typically a few orders of magnitude higher than that of dark count rates. Dark counts will then only sightly add to the error rate. In our scheme, memory decoherence errors play a major role as we will explain in this and the following sections.

In the following, we describe the protocol and its components in more detail.

\subsection{Protocol}

In our protocol, Alice and Bob, at a rate $R_S$, send BB84 encoded pulses to the middle station (dashed boxes in figure~\ref{Fig:figure1_a_b_c}). At the QMs, for each incoming pulse, we either apply a loading process by which we can store the state of the photons into memories and verify it, or use the indirectly heralding scheme of figure~\ref{Fig:figure1_a_b_c}(b). Once successful for a particular QM, we stop the loading procedure on that QM, and wait until both memories are loaded, at which point, a BSM is performed on the QMs. The BSM results are sent back to Alice and Bob, and the above procedure is being repeated until a sufficient number of raw key bits is obtained. The rest of the protocol is the same as that of MDI-QKD. Sifting and postprocessing will be performed on the raw key to obtain a secret key. In this paper, we neglect the finite-size-key effects in our analysis \cite{MDIQKD_finite_PhysRevA2012}.


\subsection{Component modeling}

In this section, we model each component of figure (\ref{Fig:figure1_a_b_c}) including sources and encoders, the channel, QMs, and the BSM module. 

\subsubsection{Sources and encoders}

We consider two types of sources: ideal single-photon sources and phase-randomized weak laser pulses. The latter will be used in the decoy-state \cite{Lo:Decoy:2005} version of the protocol. Each source, at both Alice's and Bob's sides, generates pulses at a rate $R_S$. Each pulse is polarization encoded in either the rectilinear ($Z$) or diagonal ($X$) basis. {In the case of ideal single photons, we, correspondingly, send states $|H\rangle$ and $|V\rangle$ in the $Z$ basis, and $(|H\rangle + |V\rangle)/\sqrt{2}$ and $(|H\rangle - |V\rangle)/\sqrt{2}$ in the $X$ basis. In each basis, the two employed states, respectively, represent bits 1 and 0. In the case of the decoy-state protocol, the single-photon states are replaced with weak phase-randomized coherent states of the same polarization.} Here, we use the efficient version of BB84 encoding, where the $Z$ basis is used much more frequently than the $X$ basis \cite{Lo:EffBB84:2005}. {The pulse duration is denoted by $\tau_p$ and it is chosen in accordance with the requirements of the memory system in use.}

There are several sources of nonideality one may be concerned with at the encoder box. For instance, in \cite{Chan:MIQKDexp:2013}, one major source of error is in not generating fully orthogonal states in each basis. {Note that secure exchange of keys may still be possible, although at a possibly reduced rate, by using even uncharacterised sources \cite{Han:uncharSource_PRA2013}.} Another possible issue would be in having multiple-photon components if one uses parametric down-converters to generate single photons \cite{Razavi_CohMeas_PRA09,Razavi_CohMeas_JPhysB09}. Although all these issues, among others, are important in the overall performance of the system, here we would rather focus on the memory side of the system, which is newly introduced, and deal with the details of source imperfections, and their effects on the secret key generation rate in a separate publication \cite{LoPiparo:2014}.

\subsubsection{Channels}

The distance between Alice (Bob) and the respective QM is denoted by $L_A$ ($L_B$). The total distance between Alice and Bob is denoted by $L=L_A +L_B$. The transmission coefficient for a channel with length $l$ is given by 
\begin{equation}
\eta_{\rm ch}(l) \equiv \exp{(-l/L_{\rm att})},
\end{equation}
where $L_{\rm att}$ is the attenuation length of the channel (roughly, 22~km for 0.2 dB/km of loss).

The channel is considered to have a background rate of $\gamma_{\rm BG}$ per polarization mode, which results in an average $p_{\rm BG} = 2 \gamma_{\rm BG} \tau_p$ background photons per pulse. This can stem from stray light or crosstalk from other channels, especially if classical signals are multiplexed with quantum ones in a network setup \cite{Shields.PRX.coexist, Townsend_QI_home_2011, Telcordia_1550_1550,Telcordia_1550_1310,Razavi_MulipleAccessQKD}.

We also consider setup misalignment in our analysis. We assume certain polarization maintenance schemes are in place for the Alice's and Bob's channels, so that the reference frames at the sources and memories are, on average, the same. We, nevertheless, consider a setup misalignment error probability $e_{dK}$, for $K=A,B$, to represent misalignment errors in each channel.
 
\subsubsection{Quantum memories}


We use the following assumptions and terminologies for the employed QMs. This list covers most relevant parameters in an experimental setup {relying on polarization encoding}, whether the QM is operated in the directly or indirectly heralding mode.
\begin{itemize}
\item{In the case of a successful loading, each QM in figure~\ref{Fig:figure1_a_b_c} ideally stores a polarization {\em qubit} corresponding to the polarization of the incoming pulse. We assume that {such a squashing operation occurs \cite{BML_Squash_08, Fung:2011:Squash}} even if at the input of the QM there is a non-qubit state, e.g., a {phase-randomized coherent state. That is, if, for instance, two photons with horizontal polarizations are at the input of the memory, the QM would only store the polarization information, and ignores the photon-number information. In practice, the loading efficiency would be a function of input photon numbers, but, for simplicity, here we neglect this dependence. This is in line with our single-excitation assumption we have adopted in this paper. }
One suitable energy level structure for such a memory is the double-$\Lambda$ configuration in figure~\ref{Fig:figure1_a_b_c}(d), with a common ground state and two other metastable states corresponding to two orthogonal polarizations. The excited states can then facilitate Raman transitions from the ground state to each of the metastable states, using known optical transition techniques \cite{STIRAP:PhysRevA.1989, Razavi.Memory.07}, in response to the input polarization state.

We assume that each QM only stores one spatio-temporal mode of light. Our protocol can be extended to incorporate multimode QMs \cite{Multimode_Gisin_PRL07, Gisin_Nature_AFC_2011, Tittel_Nature_AFC_2011, Tittel:Ondemand_2013} or multiple QMs \cite{Razavi.Lutkenhaus.09}, in which case a linear improvement in the rate is expected. In this work, we focus on the case of a single logical memory per user and leave extensions to future work.}
\item{For directly heralding memories, we denote the QM's writing efficiency by $\eta_w$. The writing efficiency is the probability to store a qubit and herald success conditioned on having a single-photon at the QM's input. Note that $\eta_w$ also includes the chance of failure for our verification process. For indirectly heralding memories, we introduce an entangling efficiency, $\eta_{\rm ent}$, which is the probability of success for entangling a photon with our QM.}
\item{We denote the QM's reading efficiency by $\eta_r$. That is the probability to retrieve a single photon out of the QM conditioned on a successful loading in the past. The reading efficiency is expected to decay over a time period $t$ as $\eta_r(t) = \eta_{r0} \exp{[-t/T_1]}$, where $T_1$ is the memory amplitude decay time and $\eta_{r0}$ is the reading efficiency right after loading. {In our example of a double-$\Lambda$-level memory of figure~\ref{Fig:figure1_a_b_c}(d), such a decay corresponds to the transition form one of the metastable states $|s_H\rangle$ or $|s_V\rangle$ to the ground state $|g\rangle$, in which case, no photon will be retrieved from the memory.}} 
\item{We denote the QM's writing time by $\tau_w$. For directly heralding memories, it is the time difference between the time that a pulse arrives (beginning of the pulse) at the QM and the time that a successful/unsuccessful loading is declared. This is practically the fastest repeat period one can run our protocol. In the case of indirectly heralding memories, $\tau_w$ includes the time for the entangling process as well as that of the side BSM operation. Accounting for such timing parameters is essential in enabling us to have a fair comparison between memory-assisted and no-memory QKD systems.

{One must note that in a practical setup there will be time periods, e.g., for synchronization purposes or memory refreshing, over which no raw key is exchanged. The total number of key bits exchanged over a period of time must therefore exclude such periods once the total key generation rate is calculated. In our work, we neglect all these overhead times, with the understanding that one can easily modify our final result by considering the percentage of the time spent on such processes within a specific practical setup.}}
\item{We denote the QM's reading time by $\tau_r$. It is the time difference between the time that the retrieval process is applied until a pulse (end of the pulse) is out.}
\item{We denote the QM's coherence (dephasing) time by $T_2$. For an initial state $\rho(0)$ of the QM at time zero, its state at a later time $t$ is given by \cite{Razavi.Lutkenhaus.09}
\begin{equation}
\label{dephasing}
\rho(t) = p(t) \rho(0) + [1-p(t)] Z \rho(0) Z,
\end{equation}
where
$p(t) = [1+ \exp(-t/T_2)]/2$. Note that dephasing would only occur if we are in a superposition of $Z$ eigenstates, e.g., the eigenstates of $X$. {The above model of decoherence is expected to have more relevance in some practical cases of interest \cite{MIT-NU, Rempe:Nature:2012, Razavi.Lutkenhaus.09} than }the model used in \cite{Brus:MDIQKD-QM_2013}, in which the memory state switches suddenly from an intact one to a fully randomized version after a certain time. We discuss the implications of each model in our numerical result section.
{It is, however, beyond the scope of this paper to fully model every possible decoherence mechanisms in QMs. Specific adjustments are needed if one uses a memory that is not properly modelled by our $T_1$ and $T_2$ time constants.}}
\end{itemize}

\subsubsection{BSM module}

\begin{figure}[hbt]
\centering
{\includegraphics[width = 7.5cm]{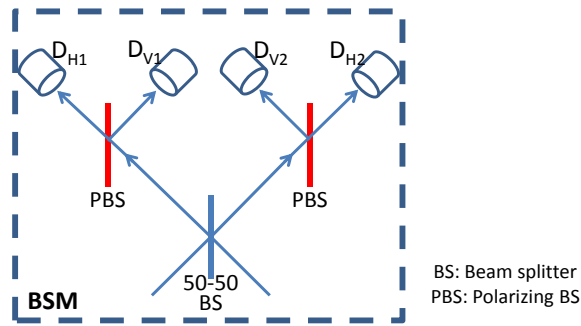}}
\caption{Bell-state measurement module for polarization states.}
\label{fig:figure_2}
\end{figure}

Figure~\ref{fig:figure_2} shows the schematic of the BSM module used in our analysis. This module enables an incomplete BSM over photonic states. In order to use this module, in our scheme, we first need to read out the QMs and convert their qubit states into polarization-encoded photons. The BSM will then be successful if exactly two detectors click, one $H$-labelled and one $V$-labelled. Depending on which detectors have clicked and what basis is in use, Alice and Bob can identify what bits they ideally share \cite{MXF:MIQKD:2012}. 

We assume the BSM module is symmetric. We lump detector quantum efficiencies with other possible sources of loss in the BSM module and denote it by $\eta_d$ for each detector. {We also assume that each detector has a dark count rate of $\gamma_{\rm dc}$, which results in a probability $p_{\rm dc} = \gamma_{\rm dc} \tau_p$ of having a dark count per pulse. The implicit assumption here is that the retrieved and the writing photons have the same pulse width. Finally, we assume that there is no additional misalignment error in the BSM module.} 

\section{Key Rate Analysis}
In this section, we find the secret key generation rate for our proposed schemes in figures~\ref{Fig:figure1_a_b_c}(a) and \ref{Fig:figure1_a_b_c}(b). {The common assumption in our predicting the relevant observed parameters in a QKD experiment is that we work under the normal mode of operation, where there is no eavesdropper present, and we are only affected by the imperfections of the system, behind which an eavesdropper can in principle hide.} We later compare our results with two conventional QKD schemes, namely, BB84, summarized in Appendix A,  and the original MDI-QKD in figure~\ref{Fig:figure1_a_b_c}(c), summarized in Appendix B, that use no memories. In all cases, we consider both single-photon and decoy-state sources. In all forthcoming sections, $f$ denotes the inefficiency of the error correction scheme, {i.e., the ratio between the actual cost of error correction and its minimum value obtained by the Shannon's theorem}, assumed to be constant, and we denote the binary entropy function as $h(p)=-p\log_2(p)-(1-p)\log_2(1-p)$, for $0\leq p\leq 1$. 
\subsection{Key rate for single-photon sources}
With ideal single-photon sources, the secret key generation rate in the setups of figures~\ref{Fig:figure1_a_b_c}(a) and \ref{Fig:figure1_a_b_c}(b) is lower bounded by \cite{ShorPreskill_00}
\begin{equation}
\label{RQM1}
R_{\rm QM} = R_S Y_{11}^{\rm QM} [1- h(e_{11;X}^{\rm QM}) - f h(e_{11;Z}^{\rm QM})],
\end{equation}
where efficient BB84 encoding is employed \cite{Lo:EffBB84:2005}. In the above equation, $e_{11;X}^{\rm QM}$ and $e_{11;Z}^{\rm QM}$, respectively, represent the quantum bit error rate (QBER) between Alice and Bob in the $X$ and $Z$ basis, when single photons are used, and $Y_{11}^{\rm QM}$ represents the probability that both memories are loaded with single photons of the same basis {\em and} the middle BSM is successful. 

To obtain the individual terms in equation (\eqref{RQM1}), we can decompose the protocol into two parts: the memory loading step and the measurement step, once both memories are loaded. The first step is a probabilistic problem with two geometric random variables, $N_A$ and $N_B$, corresponding, respectively, to the number of attempts until we load Alice and Bob's memories with single photons. The number of rounds that it takes to load both memories is then $\max\{N_A,N_B\}$. Once both memories are loaded, the rest of the protocol is similar to that of original MDI-QKD in terms of rate analysis: the QMs replace the sources in figure~\ref{Fig:figure1_a_b_c}(c) and the total transmission-detection efficiency is replaced by the reading-measurement efficiency in the BSM module. We can therefore use many of the relationships obtained for the original MDI-QKD, summarized in \ref{App:MDIQKD}, for the memory-assisted versions of figure~\ref{Fig:figure1_a_b_c}.



{For finite values of $T_1$, the reading efficiency for the Alice's QM could be different from that of Bob. In fact, we can assume that, once both memories are loaded, one of the memories (late) will be read immediately, while the other (early) $|N_A-N_B|$ rounds after its successful loading. The effective measurement efficiency for the leg $K$, $K=A,B$, corresponding to the path originating from memory $K$ in the BSM module will then be given by
\begin{equation}
\eta_{mK} = \left\{\begin{array}{cl} 
\eta_m \equiv \eta_{r0} \eta_d,& \mbox{if memory $K$ is late} \\
\eta_d \, \eta_r(t=|N_A-N_B|T),& \mbox{if memory $K$ is early}
\end{array}
\right. .
\end{equation}


With the above setting, and considering the required time for reading from the QMs, we obtain
\begin{eqnarray}
\label{Y11QM}
Y_{11}^{\rm QM} &=& \frac{1}{N_L(\eta_{1A},\eta_{1B}) + N_{r}} {\rm E}\{Y_{11}(\eta_{mA},\eta_{mB})\} \nonumber \\ 
&=& \frac{1}{N_L(\eta_{1A},\eta_{1B}) + N_{r}} Y_{11}(\eta_m,\eta'_m),
\end{eqnarray}
where $Y_{11}$ is the corresponding yield term, given by equation~\eqref{MIFluc:Model:eY11sim}, for the MDI-QKD protocol and $N_L = {\rm E}\{\max(N_A,N_B)\}$ is given by equation~\eqref{Emax}. Here, ${\rm E}\{\cdot\}$ represents the expectation value operator with respect to $N_A$ and $N_B$, and $\eta'_m = \eta_d \overline{\eta_{r}}$, where $\overline{\eta_{r}} = \eta_{r0} {\rm E}\{\exp(-|N_A-N_B|T/T_1)\}$ can be obtained from equation~\eqref{dephtime}. In equation~\eqref{Y11QM}, $N_{r}$, represents the extra rounds lost due to the nonzero reading times of QMs, once they are both loaded, and is given by
\begin{equation}
N_{r} = \left\lceil \frac{\tau_r +\tau_w}{T} \right\rceil - 1, \mbox{\ \ \ \ }\tau_r,\tau_w > 0,\mbox{\ \ \ } \tau_w \leq T ,
\end{equation}
where $T=1/R_S$ is the repetition period. The condition $\tau_w \leq T$ is a matter of practicality as sending photons faster than they can be stored is of no benefit. The fastest possible rate is then obtained at $T=\tau_w$.

In the case of directly heralding memories of figure~\ref{Fig:figure1_a_b_c}(a), we have
\begin{eqnarray}
&\eta_{1K} = 1-(1-\eta_w\eta_{\rm ch}(L_K)) e^{-\eta_w p_{\rm BG}}, \quad \mbox{$K=A,B,$}&
\end{eqnarray}
as the probabilities of successful loading of Alice and Bob's QMs with single-photon sources (or background noise). {In the case of indirectly heralding memories of figure~\ref{Fig:figure1_a_b_c}(b), following our discussion in Section (\ref{Sec:System}) about the equivalence of each leg of figure~\ref{Fig:figure1_a_b_c}(b) to an asymmetric MDI-QKD system, we have
\begin{eqnarray}
&\eta_{1K} = Y_{11}(\eta_d \eta_{\rm ch}(L_K), \eta_d \eta_{\rm ent}), \quad \mbox{$K=A,B$,}&
\end{eqnarray}
where the above terms must be calculated at an effective dark count rate of $\gamma_{\rm dc} + \gamma_{\rm BG}\eta_d/2$. }

We remark that, although obtained from different methods, the analysis in \cite{Brus:MDIQKD-QM_2013} also finds similar expressions for the yield term. In \cite{Brus:MDIQKD-QM_2013}, the analysis is only concerned with the symmetric setup, and some of the parameters considered in our work take their ideal values. It can be verified, however, that in the special case of $\tau_w = T$, $\tau_r =0$, $\gamma_{\rm BG} = 0$, $L_A = L_B$, $\eta_w =1$, and $T_1 \rightarrow \infty$, for directly heralding memories, equation~\eqref{Y11QM} reduces to the same result obtained in \cite{Brus:MDIQKD-QM_2013}. By accounting for additional relevant parameters, our analysis offers a better match to realistic experimental scenarios. 


Similarly, the error terms are given by
\begin{eqnarray}
\label{e11new}
e_{11;Z}^{\rm QM} &=& {\rm E}\{ e_{11;Z}(\eta_{mA},\eta_{mB},e_{dZ}^{\rm QM}(\eta_{1A},\eta_{1B})) \}\nonumber \\
&=&e_{11;Z}(\eta_m,\eta'_m,e_{dZ}^{\rm QM}(\eta_{1A},\eta_{1B})), \nonumber \\
e_{11;X}^{\rm QM} &=& {\rm E}\{e_{11;X}(\eta_{mA},\eta_{mB},e_{dX}^{\rm QM}(\eta_{1A},\eta_{1B})) \}  \nonumber \\
&\approx& e_{11;X}(\eta_m,\eta'_m, {\rm E}\{e_{dX}^{\rm QM}(\eta_{1A},\eta_{1B})\}),	
\end{eqnarray}
where, $e_{11;Z}$ and $e_{11;X}$, given by equation~\eqref{MIFluc:Model:eY11sim}, are the corresponding error terms for the original MDI-QKD. In addition to the typical sources of error, such as loss and dark count, the above expressions are functions of misalignment parameters. This misalignment could be a statistical error in the polarization stability of our setup, modelled by $e_{dA}$ and $e_{dB}$, or an effective misalignment because of memory dephasing \cite{LoPiparo:2013} and/or background photons. {Putting all these effects together, as we have done in \ref{App:misalign}, we obtain 
\begin{eqnarray}
\label{edSQM}
e_{dS}^{\rm QM} (\eta_{A},\eta_{B}) &=& e_{dS}^{(A)} (\eta_{A},\eta_{B})(1-e_{dS}^{(B)} (\eta_{A},\eta_{B})) \nonumber \\
&+& e_{dS}^{(B)} (\eta_{A},\eta_{B})(1-e_{dS}^{(A)} (\eta_{A},\eta_{B})) , \quad\mbox{$S=X,Z$},
\end{eqnarray}
where $e_{dS}^{(A)}$ and $e_{dS}^{(B)}$, respectively, represent the misalignment probabilities for Alice's and Bob's memories, for basis $S=X,Z$, at loading probabilities $\eta_A$ and $\eta_B$ and are given by equations \eqref{misalAB} and \eqref{edXA}. The above equation accounts for the fact that if the state of both memories are flipped, Alice and Bob will still share identical key bits. We assume that the BSM module is balanced and does not have any setup misalignment. 

{Note that in equation~\eqref{edSQM}, because of no dephasing errors for the $Z$ eigenstates, $e_{dZ}^{\rm QM}$ is independent of $N_A$ and $N_B$, whereas $e_{dX}^{\rm QM}$ is a function of them. The approximation in equation~\eqref{e11new} assumes ${\rm E}\{e_{dX}^{\rm QM} \eta_{mA} \eta_{mB} \} \approx {\rm E}\{e_{dX}^{\rm QM} \} {\rm E}\{\eta_{mA} \eta_{mB} \}$, which is valid when $T_1 \gg T_2$, to give a more readable final result.}

Equation \eqref{edSQM} can also be used in the case of indirectly heralding QMs as explained in \ref{App:misalign}. The main idea is to use the analogy of each leg in figure~\ref{Fig:figure1_a_b_c}(b) with the original MDI-QKD in figure~\ref{Fig:figure1_a_b_c}(c).}


 
\subsection{Key rate for decoy states}
Suppose Alice and Bob use a decoy-state scheme with average photon numbers $\mu$ and $\nu$, respectively, for the two main signal intensities, and infinitely many auxiliary decoy states. The secret key generation rate, in the limit of infinitely long key, is then given by
\begin{equation}
\label{RQM_decoy}
R_{\rm QM} = R_S [ Q_{11}^{\rm QM} (1- h(e_{11;X}^{\rm QM})) - f Q_{\mu\nu;Z}^{\rm QM} h(E_{\mu \nu;Z}^{\rm QM})],
\end{equation}
where
\begin{equation}
\label{QmnQM}
Q_{\mu \nu;Z}^{\rm QM} = \frac{1}{N_L(\eta_{\mu A},\eta_{\nu B}) + N_{r}} Y_{11}(\eta_m,\eta'_m)
\end{equation}
is the rate at which both memories are loaded, by Alice (Bob) sending a coherent state in the $Z$ basis with $\mu$ ($\nu$) average number of photons, and a successful BSM is achieved. In the case of directly heralding memories,  
\begin{equation}
\eta_{\mu A} = 1-e^{-\eta_{\rm ch}(L_A) \eta_w \mu - \eta_wp_{\rm BG}} \mbox{\ and \ } 
\eta_{\mu B} = 1-e^{-\eta_{\rm ch}(L_B) \eta_w \nu - \eta_wp_{\rm BG}}
\end{equation}
are the probabilities for successful loading of Alice and Bob's QMs with coherent-state sources. Similarly,
\begin{equation}
E_{\mu \nu;Z}^{\rm QM} = e_{11;Z}(\eta_m,\eta'_m,e_{dZ}^{\rm QM}(\eta_{\mu A},\eta_{\nu B}))
\end{equation}
is the QBER in the $Z$ basis, and
\begin{equation}
Q_{11}^{\rm QM} = Q_{\mu \nu;Z}^{\rm QM}\frac{\eta_{1A}\eta_{1B}}{\eta_{\mu A}\eta_{\nu B}}\mu\nu e^{-\mu-\nu}
\end{equation}
is the contribution of single-photon states in the gain term of equation~\eqref{QmnQM}.

Similar to the treatment in the previous subsection, one can find or approximate the above terms in the case of indirectly heralding memories as well. For the sake of brevity, we leave this extension to the reader.

Apart from all additional parameters considered in our model as compared to \cite{Brus:MDIQKD-QM_2013}, our treatment of the decoy-state QKD is different from that of \cite{Brus:MDIQKD-QM_2013} in the way that QMs are modelled. In our work, we assume QMs store qubits, which while is not necessarily an exact model, it often serves a good first-order approximation to the reality. In \cite{Brus:MDIQKD-QM_2013}, however, QMs are assumed to be able to store number states. This assumption seems more restrictive as many QMs, such as single trapped atoms or ions, can only store one photon.
\subsection{Storage time}
\label{Sec:sym}
To get some insight into the working of our system, in this section, we simulate the achievable rates assuming $L_A = L_B = L/2$. 
The average number of trials to load both memories from equation\eqref{Emax} is then given by \cite{Kuzmich_MultipleMem_PRL07}
\begin{equation}
N_L(\eta,\eta) = \frac{3-2 \eta}{\eta (2-\eta)} \approx \frac{3}{2}\cdot \frac{1}{\eta}, \quad\mbox{for $\eta \ll 1$,}
\end{equation}
where $\eta$ is the probability of successfully loading a QM at distance $L/2$, approximately, given by $\eta_{\rm QM} \exp(-(L/2)/L_{\rm att})$, where $\eta_{\rm QM} =\eta_w$ for directly heralding memories, and $\eta_{\rm QM} = \eta_{\rm ent} \eta_d^2$ for indirectly heralding QMs. Similarly, the average required storage time, from equation~\eqref{Tst}, is given by
\begin{equation}
T_{st} = {\rm E}\{|N_A-N_B|\} T = \frac{2(1-\eta)T}{\eta (2-\eta)} \approx \frac{T}{\eta}, \quad\mbox{for $\eta \ll 1$}, 
\end{equation}
which is similar to the result reported in \cite{Brus:MDIQKD-QM_2013}.

\begin{figure}[hbt]
\centering
\includegraphics[width = 7.5cm]{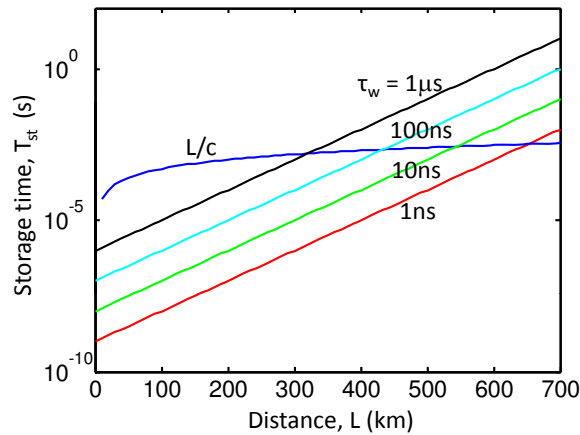}
\caption{Average required storage time, $T_{st}$, versus distance, in our scheme, for different repetition rates $1/\tau_w$. As compared to that of a probabilistic quantum repeater, labelled by $L/c$, where $c = 2 \times 10^8$~m/s is the speed of light in optical fibre, our scheme requires lower coherence times up to a certain distance. The crossover distance at $\tau_w = 1$~$\mu$s is over 300~km and at $\tau_w = 1$~ns is nearly 700~km. In all curves, $\eta_w = \eta_d = \eta_{\rm ent} = 1$ and $p_{\rm BG} =0$.}
\label{Fig:figure_3}
\end{figure}

The secret key generation rate in equations \eqref{RQM1} and \eqref{RQM_decoy} is proportional to the pulse generation rate  $R_S=1/T$ at the encoder. To maximize $R_S$, we choose $T= \tau_w$, throughout this section and next, resulting in $T_{st} \approx \tau_w/\eta$. Figure~\ref{Fig:figure_3} compares $T_{st}$ with the required storage time in multi-memory probabilistic quantum repeaters \cite{Razavi.Lutkenhaus.09}, $L/c$, where $c$ is the speed of light in the channel. It can be seen that our scheme offers lower required coherence times until a certain distance. With fast memories of shorter than 10~ns of access time, this crossover distance could be longer than 500~km. With such memories, the required coherence time at 300~km is roughly 1~$\mu$s, or lower, as compared to over 1~ms for probabilistic quantum repeaters.

It is worth mentioning that the possible advantage of requiring low coherence times is only achievable for systems with nesting level one, {i.e., with one stage of entanglement swapping}. Unlike quantum repeaters, our protocol, in terms of its timing, is not scalable to higher nesting levels. Nevertheless, even with only one entanglement swapping stage, our protocol can outperform conventional QKD schemes in terms of rate-versus-distance behaviour, and, more importantly, this can possibly be achieved with existing technology for quantum memories. We explore this and other aspects of our scheme in the next section.

\section{Numerical Results}
In this section, we study the impact of various parameters on the secret key generation rate of our scheme. All results have been obtained assuming the symmetric setup described in Section (\ref{Sec:sym}), $\tau_w=T$, $f=1.16$, $c=2 \times 10^8$~m/s, and 0.2~dB/km of loss in the channel. We also compare our scheme with the efficient BB84 and MDI-QKD protocols, whose secret key generation rates are, respectively, summarized in Appendices A and B.
\subsection{Coherence time}
\begin{figure}[hbt]
\centering
\includegraphics[width = 7.5cm]{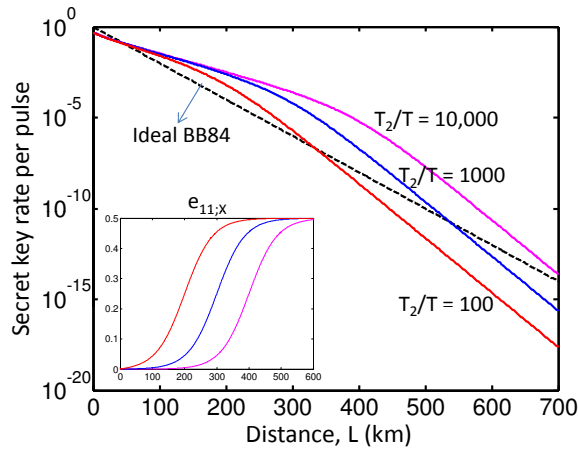}
\caption{Secret key generation rate per pulse for the heralded scheme of figure~\ref{Fig:figure1_a_b_c}(a) for different values of $T_2/T$ using single-photon sources. The dashed line represents the ideal efficient BB84 case. Unless explicitly mentioned, all other parameters assume their ideal values: $T_1 \rightarrow \infty$, $\eta_w = \eta_{r0} =\eta_d = 1$, $\gamma_{\rm BG} = \gamma_{\rm dc} =0$, $e_{dA}=e_{dB}=0$, and $\tau_r =0$.}
\label{Fig:figure_4}
\end{figure}

{In this section, we discuss the effects of memory dephasing on the secret key generation rate. As mentioned before, while our scheme in figure~\ref{Fig:figure1_a_b_c}(a) is particularly resilient to dark count errors, it still suffers from memory errors.} Figure~\ref{Fig:figure_4} demonstrates the secret key generation rate per pulse at different coherence times for the scheme of figure~\ref{Fig:figure1_a_b_c}(a). A finite coherence time is the only source of nonideality considered in this figure. Since, in our model, the dephasing process only affects the diagonal basis, $e_{11;Z}^{\rm QM} = 0$ at all distances; hence $R_{\rm QM} \propto 1- h(e_{11;X}^{\rm QM})$ remains always positive. The rate is initially proportional to $\exp(-(L/2)/L_{\rm att})$, and with low values of $e_{11;X}^{\rm QM}$ for short distances, our scheme beats the BB84 case depicted by the dashed line. {Note that, because of the partial BSM in figure~\ref{fig:figure_2}, the initial key rate at $L=0$ for our scheme is lower than that of BB84}. At large distances, however, the dephasing process becomes significant and results in $e_{11;X}^{\rm QM}$ approaching 1/2; see the inset. Subsequently, $R_{\rm QM}$ decays with a faster slope and at some point becomes lower than what one can achieve with an ideal BB84 system. The window between the two crossing points on each curve is the range where our scheme, can, in principle, beat a noise-free BB84 system. This window is larger for QMs with longer coherence times. 

\begin{figure}[hbt]
\centering
\includegraphics[width = 7.5cm]{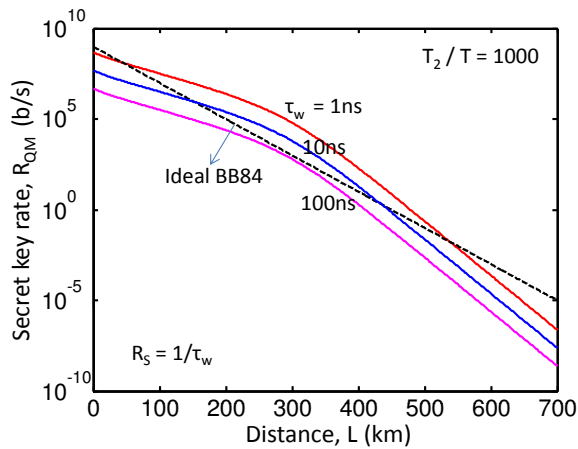}
\caption{Secret key generation rate for different values of $\tau_w$ at $T_2/T =1000$ using single-photon sources. The dashed line represents the ideal efficient BB84 case at $R_S =$~1~Gpulse/s. Unless explicitly mentioned, all other parameters assume their ideal values: $T_1 \rightarrow \infty$, $\eta_w = \eta_{r0} =\eta_d = 1$, $\gamma_{\rm BG} = \gamma_{\rm dc} =0$, $e_{dA}=e_{dB}=0$, and $\tau_r =0$.}
\label{Fig:figure_5}
\end{figure}

In \cite{Brus:MDIQKD-QM_2013}, authors look at the minimum required coherence time to achieve nonzero key rates, assuming $e_{11;X}^{\rm QM} = e_{11;Z}^{\rm QM}$ within their model of decoherence. Although the models used for decoherence in our work and \cite{Brus:MDIQKD-QM_2013} are different, $e_{11;X}^{\rm QM}$ has a similar behaviour in both cases. In our case, however, the transition from 0 to 1/2 is smoother than that of \cite{Brus:MDIQKD-QM_2013}. This is expected as the model in \cite{Brus:MDIQKD-QM_2013} is an abrupt good-bad model for the memory. A consequence of this difference is that the minimum required coherence time is then higher in our case, which highlights the importance of the more accurate model we have used for decoherence.
   
The comparison in figure~\ref{Fig:figure_4} assumes that the source rate $R_S$ is the same for both the BB84 protocol and our scheme. In our scheme, however, $R_S$ depends on the writing time of the memories. Figure~\ref{Fig:figure_5} shows the secret key generation rate for the scheme of figure~\ref{Fig:figure1_a_b_c}(a) at a fixed value of $T_2/T=1000$, but for several values of $T = \tau_w = 1/R_S$. The BB84 system is run at a fixed rate of 1~GHz. Again, we assume that the only source of nonideality is memory dephasing. It can be seen that slow memories with writing times of 100~ns, or higher, can hardly compete with an ideal BB84 system. The two orders of magnitude lost because of the lower repetition rate cannot be compensated within the first 300~km. It is still possible to beat the BB84 case, at long distances, if memories have higher coherence times.


\subsection{Realistic examples}

{It is interesting to see if any of the existing technologies for quantum devices can be employed in our scheme to beat conventional QKD systems. Figure~\ref{Fig:figure_6} makes such a comparison between BB84, MDI-QKD, and memory-assisted MDI-QKD for particular experimental parameters. We have chosen our QM parameters based on the two lessons learned from figures~\ref{Fig:figure_4} and \ref{Fig:figure_5}: the QM needs to have a high bandwidth-storage product ($T_2 / \tau_w$) on the order of 1000 or higher, and, it also needs to be fast, with writing times on the order of nanoseconds. Both these criteria are met for the QM used in \cite{Walmsley:PRL:2010}, which particularly offers fast reading and writing with 300-ps-long pulses at a storage time of around 4~$\mu$s. The employed memory in this experiment is an atomic ensemble, which fits our indirectly heralding scheme of figure~\ref{Fig:figure1_a_b_c}(b). {We should, however, be careful with multiple excitations in this case, which are not considered in our model.} We therefore assume that, by driving this memory with short pulses, one can ideally generate the jointly entangled state in equation~\eqref{jointent} between the memory and a photon \cite{Pan:NatPhys:2012}, where, in this case, $|s_H\rangle$ and $|s_V\rangle$ are, respectively, the corresponding symmetric collective excited states to horizontal and vertical polarizations \cite{DLCZ_01,Kuzmich_memory_05}. By keeping the entangling efficiency low at $\eta_{\rm ent} = 0.05$, here, we try to keep the effect of multiple excitations in such memories low \cite{Razavi.DLCZ.06, Razavi.Amirloo.10, LoPiparo:2013}; further analysis is, however, required to fully account for such effects \cite{LoPiparo:2014}. We also assume that $T_2 = T_1$ and use the state-of-the-art single-photon detectors with $\eta_d = 0.93$ at $\gamma_{\rm dc} = 1$ count per second and 150~ps of time resolution \cite{Nam_NatPhot_93p_2013} for all systems. 

\begin{figure}[hbt]
\centering
\includegraphics[width = 7.5cm]{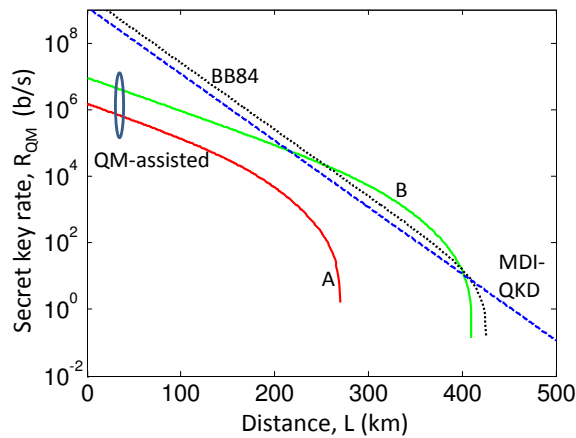}
\caption{Secret key generation rate for single-photon BB84 (dotted), MDI-QKD (dashed), and our indirectly heralding scheme of figure~\ref{Fig:figure1_a_b_c}(b) (solid) at practical parameter values. In all curves, $\eta_d = 0.93$, $\gamma_{\rm dc} = 1$/s, $\gamma_{\rm BG} = 0$, and $e_{dA}=e_{dB}=0.005$. For BB84 and MDI-QKD, $R_S=$~3.3~G~pulse/s, similar to $R_S = 1/\tau_w$, in our scheme. For our scheme, we have used some of the experimental parameters reported in \cite{Walmsley:PRL:2010}. For the curve labelled A, $\eta_{\rm ent} = 0.05$, $\eta_{r0} = 0.3$, $T_1 = T_2 = 4$~$\mu$s, and $\tau_w = \tau_r = \tau_p = 300$~ps. It is assumed that there is no multiple excitations in the QMs. For the curve labelled B, everything is the same except that $\eta_{r0} = 0.73$ and $T_1 = T_2 = 100$~$\mu$s.}
\label{Fig:figure_6}
\end{figure}

We consider two sets of parameter values for our employed QM in figure~\ref{Fig:figure_6}. In the first set, corresponding to the curve labelled A on the figure, we use the same numerical values as reported in \cite{Walmsley:PRL:2010}, that is, $\eta_{r0} = 0.3$, $T_1 = 4$~$\mu$s, and $\tau_w = \tau_r = \tau_p = 300$~ps. We, however, assume that $R_S = 1/\tau_w$, which is much faster than the repetition rate used in \cite{Walmsley:PRL:2010}. In the curve labeled B, we improve the performance by assuming $\eta_{r0} = 0.73$, which is what another group has obtained for a similar type of memory \cite{Pan:NatPhys:2012}, and $T_1 = T_2 = 100$~$\mu$s, which is attainable by improving magnetic shielding \cite{Camacho_Natphot_2009}. It can be seen that, whereas the current QM employed in \cite{Walmsley:PRL:2010} is short of beating either of no-memory systems, our slightly boosted system, in curve B, outperforms both systems at over roughly 200~km. The cut-off distance in curve B is about 400 km, which is mainly because of memory decoherence, and it can be improved by using memories with longer coherence times. This implies that with slightly improving some experimental parameters, we would be able to employ realistic QMs to improve the performance of practical quantum communication setups. We remark that the example QM chosen in figure~\ref{Fig:figure_6} is not necessarily the only option, and improved versions of other types of memories can potentially offer the same performance \cite{Rempe:Nature:2012, Tittel_Nature_AFC_2011, Blatt:Natture:2012, Gisin:AFCmem_JLum2010,  Gisin_AFC_polar_PRL_2012, Riedmatten_AFC_polar_PRL_2012, Guo_AFC_polar_PRL_2012}.}  

{What we have proposed here is an initial step toward improving the performance of QKD systems by using quantum memories. In particular, we have shown how technologically close we are to beating a direct, no-memory, QKD link in terms of the achievable rate at certain long distances. Our scheme is not, however, scalable to arbitrarily long distances. For that matter, full quantum repeaters would eventually be needed. A possible roadmap for the development of such systems would pass through probabilistic, and then deterministic, and eventually no-memory versions of quantum repeaters \cite{Munro:NatPhot:2012, Azuma:All_optical_QR_2013, Liang:NoMemRep_2013}. It is hard to make a fair comparison between all these and our scheme, as the required resources in each case are different. Some studies have nevertheless compared different repeater schemes under certain assumptions \cite{PhysRevA.87.062335, LoPiparo:2013}. It is only the future, in the end, that proves which system, and at what price, can be implemented over the course of time.}

\section{Conclusions}

By combining ideas from quantum repeaters and MDI-QKD, we proposed a QKD scheme that relied on quantum memories. While offering the same rate-versus-distance improvement that quantum repeaters promise, the coherence-time requirements for the quantum memories employed in our scheme could be less stringent than that of a general single-mode probabilistic quantum repeater system. That would provide a window of opportunity for building realistic QKD systems that beat conventional no-memory QKD schemes by only relying on existing technologies for quantum memories. In our work, we showed that how close some experimental setups would be in achieving this objective. Our protocol acts as a middle step on the roadmap to long-distance quantum communication systems.

\section*{Acknowledgments}
This research was supported in part by the European Community's Seventh Framework Programme Grant Agreement 277110, the UK Engineering and Physical Sciences Research Council Grant No.~EP/J005762/1, the National Basic Research Program of China Grants No.~2011CBA00300 and No.~2011CBA00301, the 1000 Youth Fellowship program in China, the NSERC Discovery Program, and the DARPA Quiness Program.


\appendix

\section{BB84 Key Rate analysis}
In this appendix we summarize the secret key generation rate for the efficient BB84 protocol \cite{Lo:EffBB84:2005} shown in figure~\ref{Fig:figure_7}.
In figure~\ref{Fig:figure_7}, Alice is the transmitter sending pulses in either the rectilinear or diagonal basis and Bob is the receiver, which decodes the message. They communicate through an optical channel of distance $L$.

\begin{figure}[hbt]
\centering
{\includegraphics[width = 7.5cm]{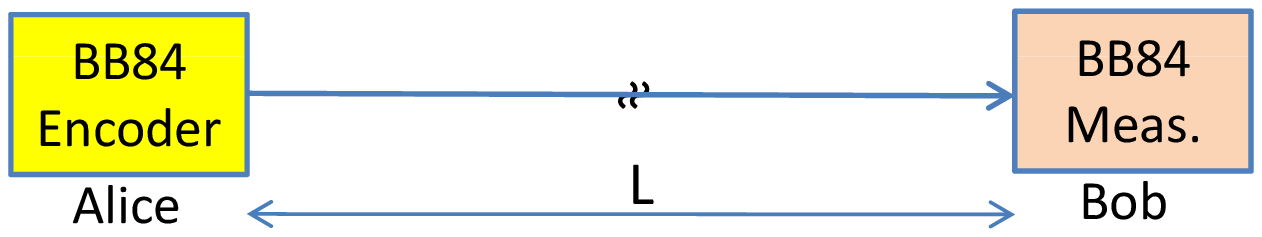}}
\caption{The setup for the BB84 protocol.}
\label{Fig:figure_7}
\end{figure}

With a clock rate of $R_S$, the secret key generation rate is lower bounded by
\begin{equation}
\label{RBB84}
R_{\rm BB84} = R_S Y_1 [1- h(e_1) - f h(e_1)],
\end{equation}
in the single-photon case, and 
\begin{equation}
\label{RBB84decoy}
R_{\rm BB84} = R_S [ Q_1 (1- h(e_1)) - f Q_\mu h(E_\mu)],
\end{equation}
in the (infinitely many) decoy-state case, where $\mu$ is the average number of photons for signal states, which is dominantly used. In equation~\eqref{RBB84}, $Y_1$ is the yield of single photons, or the probability that Bob gets a click on his measurement devices assuming that Alice has sent exactly one photon, and is given by
\begin{equation}
Y_1 = Y_C + Y_E = 1 - (1-\eta)(1- p_{\rm dc})^2,
\end{equation}
where $\eta = \eta_{\rm ch}(L) \eta_d$, and
\begin{equation} 
Y_C = (1- p_{\rm dc}/2)(\eta + (1-\eta)p_{\rm dc}) \mbox{\ and \ }
Y_E = p_{\rm dc}[(1-\eta)(1- p_{\rm dc}/2)+\eta/2]
\end{equation}
correspond, respectively, to the terms that, in the absence of misalignment, result in identical (Correct) versus non-identical (Error) bits shared by Alice and Bob. The QBER, $e_1$, which is the same for both bases, is given by
\begin{eqnarray}
e_1 Y_1 &=& e_d Y_C + (1-e_d) Y_E = e_0 Y_1 - (e_0 - e_d) (Y_C - Y_E) \nonumber \\
&=& e_0 Y_1 - (e_0 - e_d) \eta (1 - p_{\rm dc}),
\end{eqnarray}
where $e_0 =1/2$ and $e_d = e_{dA} + e_{dB}$ is the total misalignment probability for the channel.

Similarly, in equation~\eqref{RBB84decoy}, 
\begin{eqnarray}
&Q_1= Y_1 \mu e^{-\mu},& \nonumber\\
& Q_\mu = Q_C + Q_E = 1- e^{-\eta \mu}(1-p_{\rm dc})^2 ,& \nonumber \\
&Q_C = (1- p_{\rm dc}/2)(1-e^{-\eta \mu} + e^{-\eta \mu}p_{\rm dc}),& \nonumber \\
&Q_E = p_{\rm dc}[e^{-\eta \mu}(1- p_{\rm dc}/2)+(1-e^{-\eta \mu})/2],& 
\end{eqnarray}
are the corresponding gain terms \cite{Lo:Decoy:2005}, and
\begin{equation}
E_\mu Q_\mu= e_0 Q_\mu - (e_0 - e_d) (1-e^{-\eta \mu}) (1 - p_{\rm dc}), e_0 =1/2,
\end{equation} 
gives the QBER.


\section{MDI-QKD key rate analysis}
\label{App:MDIQKD}
The secret key generation rate for the MDI-QKD scheme of figure~\ref{Fig:figure1_a_b_c}(c) is lower bounded by \cite{MXF:MIQKD:2012}
\begin{equation}
R_{\rm MDI-QKD} = R_S Y_{11} [1- h(e_{11;X}) - f h(e_{11;Z})],
\end{equation}
in the single-photon case, and 
\begin{equation}
R_{\rm MDI-QKD} = R_S [ Q_{11} (1- h(e_{11;X})) - f Q_{\mu\nu;Z} h(E_{\mu \nu;Z})],
\end{equation}
in the decoy-state case, where $\mu$ ($\nu$) is the average number of photons for signal states sent by Alice (Bob). Here, $Q_{11}$ is the probability of a successful BSM if Alice and Bob, respectively, send pulses with $\mu$ and $\nu$ average number of photons and is given by
\begin{equation} \label{MIQKD:Model:Q11}
Q_{11}(\eta_a, \eta_b) = \mu\nu e^{-\mu-\nu}Y_{11}(\eta_a, \eta_b) \\
\end{equation}
where $\eta_a = \eta_{\rm ch}(L_A) \eta_d$ and $\eta_b = \eta_{\rm ch}(L_B) \eta_d$ are transmission coefficients of each leg in figure~\ref{Fig:figure1_a_b_c}(c), and \cite{MXF:MIQKD:2012}
\begin{eqnarray} \label{MIFluc:Model:eY11sim}
& Y_{11}(\eta_a, \eta_b) =(1-p_{\rm dc})^2\left[ \frac{\eta_a\eta_b}{2} +(2\eta_a+2\eta_b-3\eta_a\eta_b)p_{\rm dc} +4(1-\eta_a)(1-\eta_b)p_{\rm dc}^2 \right],& \nonumber\\
&e_{11;X}(\eta_a, \eta_b,e_d)Y_{11}(\eta_a, \eta_b) = e_0 Y_{11}(\eta_a, \eta_b)-(e_0-e_d)(1-p_{\rm dc})^2 {\eta_a\eta_b}/{2},& \nonumber\\
&e_{11;Z}(\eta_a, \eta_b,e_d)Y_{11}(\eta_a, \eta_b) = e_0 Y_{11}(\eta_a, \eta_b)-(e_0-e_d)(1-p_{\rm dc})^2 (1-2p_{\rm dc}) {\eta_a\eta_b}/{2}&\nonumber\\
&
\end{eqnarray}
with $e_d$ being the total misalignment probability. In the scheme of figure~\ref{Fig:figure1_a_b_c}(c), $e_d = e_{dA}(1-e_{dB}) + e_{dB}(1-e_{dA})$.
Similarly, using the results obtained in \cite{MXF:MIQKD:2012}, we have
\begin{eqnarray}
\label{Path:Decoy:OriGain}
&Q_{\mu\nu;Z} = Q'_C + Q'_E & \nonumber \\
&E_{\mu\nu;Z} Q_{\mu\nu;Z} = e_d Q'_C + (1-e_d) Q'_E,&
\end{eqnarray}
where
\begin{eqnarray}
&Q'_C = 2(1-p_{\rm dc})^2e^{-{\mu'}/{2}} \left[1-(1-p_{\rm dc})e^{-\eta_a\mu_a/2}\right]\left[1-(1-p_{\rm dc})e^{-\eta_b\mu_b/2}\right]& \nonumber \\
&Q'_E = 2 p_{\rm dc} (1-p_{\rm dc})^2 e^{-{\mu'}/{2}} [I_0(2x)-(1-p_{\rm dc}) e^{-{\mu'}/{2}}].&
\end{eqnarray}
In above equations, $I_0(x)$ is the modified Bessel function of the first kind and

\begin{eqnarray}\label{Path:Model:CoherentNotations}
x &= \sqrt{\eta_a\mu\eta_b\nu}/2, \\
y &= (1-p_{\rm dc})e^{-\frac14(\eta_a\mu+\eta_b\nu)}, \\
\mu' &= \eta_a \mu+\eta_b\nu.&
\end{eqnarray}


\section{Loading process}
\label{App:QM}
The loading process in the setups of figures~\ref{Fig:figure1_a_b_c}(a) and \ref{Fig:figure1_a_b_c}(b) are probabilistic ones, with two geometric random variables $N_A$ and $N_B$ playing the major role. Suppose the success probability for each loading attempt corresponding to these random variables is, respectively, given by $\eta_A$ and $\eta_B$. Then, we obtain the following probability distribution for $|N_A-N_B|$:
\begin{equation}
\Pr(|N_A - N_B| = k) = [(1-\eta_A)^k +(1- \eta_B)^k]P_0, \mbox{\ \ \ } k>0,
\end{equation} 
where
\begin{equation}
P_0 = \Pr(N_A = N_B) = \frac{\eta_A \eta_B}{\eta_A + \eta_B -\eta_A \eta_B}.
\end{equation} 
Using the above expressions, we then obtain
\begin{eqnarray}
\label{Emax}
N_L(\eta_A,\eta_B) &=& {\rm E}\{\max(N_A,N_B)\} \nonumber \\
&=& \frac{1}{2} {\rm E}\{|N_A-N_B| + N_A+N_B\} \nonumber \\
&=& \frac{1}{2} \left[ \frac{\eta_A(1-\eta_B)}{\eta_B(\eta_A + \eta_B -\eta_A \eta_B)} + \frac{\eta_B(1-\eta_A)}{\eta_A(\eta_A + \eta_B -\eta_A \eta_B)} + \frac{1}{\eta_A} + \frac{1}{\eta_B} \right].\nonumber\\
&
\end{eqnarray}
Moreover,
\begin{equation}
\label{dephtime}
{\rm E}\{\exp(-|N_A-N_B|\delta)\} = P_0 \left[ \frac{1}{1-e^{-\delta}(1-\eta_A)} + \frac{1}{1-e^{-\delta}(1-\eta_B)} - 1 \right]
\end{equation}
and the average storage time, $T_{st}$, is given by
\begin{equation}
\label{Tst}
T_{st} = {\rm E}\{|N_A-N_B|\} T= \frac{\eta_A(1-\eta_B)T}{\eta_B(\eta_A + \eta_B -\eta_A \eta_B)} + \frac{\eta_B(1-\eta_A)T}{\eta_A(\eta_A + \eta_B -\eta_A \eta_B)}. 
\end{equation}
{Finally, we can show that
\begin{equation}
\label{NANB}
\Pr\{N_A \geq N_B\} = \frac{\eta_B}{1-(1-\eta_A)(1-\eta_B)} = 1- \Pr\{N_A < N_B\}
\end{equation}
and
\begin{eqnarray}
\label{SAB}
S_{A<B}(\delta) &\equiv&  \sum_{1=n_a<n_b}^{\infty}{\Pr\{N_A = n_a, N_B = n_b\}\exp[(n_a - n_b)\delta]} \nonumber \\
&=& \frac{\eta_A \eta_B (1-\eta_B) e^{-\delta}}{[1-(1-\eta_B)e^{-\delta}][1-(1-\eta_A)(1-\eta_B)]}.
\end{eqnarray}

\section{Misalignment Parameters}
\label{App:misalign}
In this Appendix, we obtain the misalignment probability for each of the setups in figures~\ref{Fig:figure1_a_b_c}(a) and \ref{Fig:figure1_a_b_c}(b). Let us first consider the directly heralding memory case in the $Z$ basis and assume loading probabilities $\eta_A$ and $\eta_B$ for Alice's and Bob's memories. Suppose the legitimate state is $|s_H\rangle \langle s_H|$. Assuming setup misalignment probabilities $e_{dK}$, $K=A,B$, for leg $K$ of figure~\ref{Fig:figure1_a_b_c}(a), in the absence of background counts, the stored state in memory $K$ will become $\rho_{d0} = (1-e_{dK})|s_H\rangle \langle s_H| + e_{dK}|s_V\rangle \langle s_V|$. Now, including the background counts, the memory state will become
\begin{equation}
\rho_{dZ} = [1-e_{\rm BG}^{(K)}]\rho_{d0} + e_{\rm BG}^{(K)} \frac{|s_H\rangle \langle s_H|+|s_V\rangle \langle s_V|}{2},
\end{equation}
where $e_{\rm BG}^{(K)} = \frac{1-e^{- \eta_w p_{\rm BG}}}{\eta_K}$, $K=A,B$, is the probability that our memory has been loaded by a background (unpolarized) photon conditioned on a successful loading. The total misalignment probability in the $Z$ basis for the Alice's and Bob's memory is then given by
\begin{equation}
\label{misalAB}
e_{dZ}^{(K)} = e_{dK}(1-e_{\rm BG}^{(K)}) + e_{\rm BG}^{(K)}/2, \quad\mbox{$K=A,B$, for directly heralding QMs.}
\end{equation}
Now, let's assume the legitimate state, in the $X$ basis, is $|s_+\rangle \langle s_+|$, where $|s_\pm\rangle = (|s_H\rangle \pm |s_V\rangle)/\sqrt{2}$. Right after a successful loading, the state of the memory is then given by  
\begin{equation}
\rho_{dX}(0) = [1-e_{\rm BG}^{(K)}]\rho'_{d0} + e_{\rm BG}^{(K)} \frac{|s_H\rangle \langle s_H|+|s_V\rangle \langle s_V|}{2}
\end{equation}
where $\rho'_{d0}= (1-e_{dK})|s_+\rangle \langle s_+| + e_{dK}|s_-\rangle \langle s_-|$. If memory $A$ is the late memory, i.e., if $N_A \geq N_B$, then there will be no dephasing errors, in which case, $e_{dX}^{(A)} = e_{dZ}^{(A)}$. If it is the early memory, however, the dephasing operation in equation~\eqref{dephasing} will act on $\rho_{dX}(0)$ to give us
\begin{equation}
\rho_{dX}(t) = [1-e_{\rm BG}^{(K)}]\rho'_{d0}(t) + e_{\rm BG}^{(K)} \frac{|s_H\rangle \langle s_H|+|s_V\rangle \langle s_V|}{2},
\end{equation}
where $\rho'_{d0}(t)= [(1-e_{dK}) p(t) + e_{dK} (1-p(t))]|s_+\rangle \langle s_+| + [e_{dK} p(t) + (1-e_{dK}) (1- p(t))]|s_-\rangle \langle s_-|$. The misalignment probability is then given by
\begin{equation}
\label{edXA}
e_{dX}^{(K)} = e_{dZ}^{(K)} + \beta_A e_{\rm deph}^{(K)},
\end{equation}
where $\beta_K = (1-2e_{dK})(1-e_{\rm BG}^{(K)})$, $K=A,B$, and
\begin{equation}
\label{edQMA}
e_{\rm deph}^{(A)} = \left\{\begin{array}{cl}0 & N_A\geq N_B\\
(1/2) [1-\exp(-|N_A-N_B|T/T_2)] & N_A < N_B 
\end{array}
, \right. 
\end{equation}
where $N_A$ and $N_B$ are geometric random variables with success probabilities $\eta_A$ and $\eta_B$. By averaging over these variables, we obtain
\begin{equation}
{\rm E}\{e_{dX}^{(A)}\} = e_{dZ}^{(A)} + \beta_A {\rm E}\{e_{\rm deph}^{(A)}\},
\end{equation}
where
\begin{equation}
\label{edephA}
{\rm E}\{e_{\rm deph}^{(A)}\} = [\Pr\{N_A < N_B\} - S_{A<B}(T/T_2)]/2,
\end{equation}
which can be obtained from equations \eqref{NANB} and \eqref{SAB}. 
One can obtain similar expressions for $e_{dX}^{(B)}$ by swapping $A$ and $B$ in equations \eqref{edQMA}--\eqref{edephA}. 

To calculate ${\rm E}\{e_{dX}^{\rm QM}\}$ from equation~\eqref{edSQM}, the final remaining term is given by
\begin{eqnarray}
\label{edXAedXB}
{\rm E}\{e_{dX}^{(A)}e_{dX}^{(B)}\} = e_{dZ}^{(A)}e_{dZ}^{(B)} + \beta_A {\rm E}\{e_{\rm deph}^{(A)}\} e_{dZ}^{(B)} + \beta_B {\rm E}\{e_{\rm deph}^{(B)}\} e_{dZ}^{(A)},
\end{eqnarray}
where we used the fact that $e_{\rm deph}^{(A)}e_{\rm deph}^{(B)} = 0$, as one of the two terms is always zero regardless of the values of $N_A$ and $N_B$.

In the case of indirectly heralding QMs, we assume that each erroneous click on the side BSMs will effectively result in a flip to the corresponding QM state, and can also be modeled as misalignment. This assumption is valid at low distances where majority of errors are caused by the setup misalignment. We then obtain
\begin{equation}
\label{edZKindirect}
e_{dZ}^{(K)} = e_{11;Z}(\eta_d \eta_{\rm ch}(L_K),\eta_d \eta_{\rm ent},e_{dK}), \quad\mbox{$K=A,B$,}
\end{equation}
for indirectly heralding QMs, where $e_{11;Z}$ can be calculated from equation~\eqref{MIFluc:Model:eY11sim} at an equivalent dark count rate of $\gamma_{\rm dc} + \eta_d \gamma_{\rm BG}/2$. At long distances, most errors originate from dark counts or background photons, whose effective misalignment effect will approach half of $e_{11;Z}$ in the above equation. As a conservative assumption, we use the expression in equation~\eqref{edZKindirect} for all distances.

All other terms in equations \eqref{e11new} and \eqref{edSQM} can be obtained following the same expressions in eqations \eqref{edXA}--\eqref{edXAedXB} at $\beta_K=1-2 e_{dZ}^{(K)}$, for $K=A,B$, and using equation~\eqref{edZKindirect} for $e_{dZ}^{(K)}$. 
}

\section*{References}

\bibliographystyle{unsrt}

\bibliography{Bibli28Sept12}

\begin{thebibliography}{10}

\bibitem{Commercial_QKD}
See, for instance, http://www.idquantique.com.

\bibitem{Wang:260kmQKD:2012}
Shuang Wang, Wei Chen, Jun-Fu Guo, Zhen-Qiang Yin, Hong-Wei Li, Zheng Zhou,
  Guang-Can Guo, and Zheng-Fu Han.
\newblock 2 {GH}z clock quantum key distribution over 260 km of standard
  telecom fiber.
\newblock {\em Opt. Lett.}, 37(6):1008--1010, March 2012.

\bibitem{Sasaki:TokyoQKD:2011}
M.~Sasaki, M.~Fujiwara, H.~Ishizuka, W.~Klaus, K.~Wakui, M.~Takeoka, A.~Tanaka,
  K.~Yoshino, Y.~Nambu, S.~Takahashi, A.~Tajima, A.~Tomita, T.~Domeki,
  T.~Hasegawa, Y.~Sakai, H.~Kobayashi, T.~Asai, K.~Shimizu, T.~Tokura,
  T.~Tsurumaru, M.~Matsui, T.~Honjo, K.~Tamaki, H.~Takesue, Y.~Tokura, J.~F.
  Dynes, A.~R. Dixon, A.~W. Sharpe, Z.~L. Yuan, A.~J. Shields, S.~Uchikoga,
  M.~Legre, S.~Robyr, P.~Trinkler, L.~Monat, J.-B. Page, G.~Ribordy, A.~Poppe,
  A.~Allacher, O.~Maurhart, T.~Langer, M.~Peev, and A.~Zeilinger.
\newblock Field test of quantum key distribution in the {Tokyo QKD Network}.
\newblock {\em Opt. Exp.}, 19(11):10387--10409, 2011.

\bibitem{secoqc}
M.~{\rm Peev \em et al.}
\newblock The {SECOQC} quantum key distribution network in {Vienna}.
\newblock {\em New J. Phys.}, 11:075001, 2009.

\bibitem{Townsend_QI_home_2011}
I.~Choi, R.~J. Young, and P.~D. Townsend.
\newblock Quantum information to the home.
\newblock {\em New J. Phys.}, 13:063039, June 2011.

\bibitem{Shields.PRX.coexist}
K.~A. Patel, J.~F. Dynes, I.~Choi, A.~W. Sharpe, A.~R. Dixon, Z.~L. Yuan, R.~V.
  Penty, and A.~J. Shields.
\newblock Coexistence of high-bit-rate quantum key distribution and data on
  optical fiber.
\newblock {\em Phys. Rev. X}, 2:041010, Nov. 2012.

\bibitem{Tittel:expMDIQKD_PRL2013}
A.~Rubenok, J.~A. Slater, P.~Chan, I.~Lucio-Martinez, and W.~Tittel.
\newblock Real-world two-photon interference and proof-of-principle quantum key
  distribution immune to detector attacks.
\newblock {\em Phys. Rev. Lett.}, 111:130501, Sep 2013.

\bibitem{Silva_expMDIQKD_PRA2013}
T.~Ferreira~da Silva, D.~Vitoreti, G.~B. Xavier, G.~C. do~Amaral, G.~P.
  Tempor\~ao, and J.~P. von~der Weid.
\newblock Proof-of-principle demonstration of measurement-device-independent
  quantum key distribution using polarization qubits.
\newblock {\em Phys. Rev. A}, 88:052303, Nov 2013.

\bibitem{Pan_expMDIQKD_PRL2013}
Yang Liu, Teng-Yun Chen, Liu-Jun Wang, Hao Liang, Guo-Liang Shentu, Jian Wang,
  Ke~Cui, Hua-Lei Yin, Nai-Le Liu, Li~Li, Xiongfeng Ma, Jason~S. Pelc, M.~M.
  Fejer, Cheng-Zhi Peng, Qiang Zhang, and Jian-Wei Pan.
\newblock Experimental measurement-device-independent quantum key distribution.
\newblock {\em Phys. Rev. Lett.}, 111:130502, Sep 2013.

\bibitem{HKLo:expMDIQKD_2013}
Zhiyuan Tang, Zhongfa Liao, Feihu Xu, Bing Qi, Li~Qian, and Hoi-Kwong Lo.
\newblock Experimental demonstration of polarization encoding
  measurement-device-independent quantum key distribution.
\newblock {\em arXiv:1306.6134 [quant-ph]}, 2013.

\bibitem{Zoller_Qrepeater_98}
H.-J. Briegel, W.~D\"ur, J.~I. Cirac, and P.~Zoller.
\newblock Quantum repeaters: The role of imperfect local operations in quantum
  communication.
\newblock {\em Phys. Rev. Lett.}, 81(26):5932--5935, Dec. 1998.

\bibitem{Munro:NatPhot:2012}
W.~J. Munro, A.~M. Stephens, S.~J. Devitt, K.~A. Harrison, and Kae Nemoto.
\newblock Quantum communication without the necessity of quantum memories.
\newblock {\em Nat. Photon.}, 6:771--781, Oct. 2012.

\bibitem{Azuma:All_optical_QR_2013}
Koji Azuma, Kiyoshi Tamaki, and Hoi-Kwong Lo.
\newblock All photonic quantum repeaters.
\newblock {\em arXiv:1309.7207 [quant-ph]}, 2013.

\bibitem{Liang:NoMemRep_2013}
Sreraman Muralidharan, Jungsang Kim, Norbert L\"utkenhaus, Mikhail~D. Lukin,
  and Liang Jiang.
\newblock Ultrafast and fault-tolerant quantum communication across long
  distances.
\newblock {\em arXiv:1310.5291 [quant-ph]}, 2013.

\bibitem{DLCZ_01}
L.-M. Duan, M.~D. Lukin, J.~I. Cirac, and P.~Zoller.
\newblock Long-distance quantum communication with atomic ensembles and linear
  optics.
\newblock {\em Nature}, 414:413 -- 418, 2001.

\bibitem{Panayi_ICQNM12}
C.~Panayi and M.~Razavi.
\newblock Measurement device independent quantum key distribution with
  imperfect quantum memories.
\newblock In {\em Tech. Digest, The Sixth International Conference on Quantum,
  Nano and Micro Technologies}, Rome, Italy, 2012.

\bibitem{Brus:MDIQKD-QM_2013}
Silvestre Abruzzo, Hermann Kampermann, and Dagmar Bru\ss{}.
\newblock Measurement-device-independent quantum key distribution with quantum
  memories.
\newblock {\em Phys. Rev. A}, 89:012301, Jan 2014.

\bibitem{Lo:MIQKD:2012}
Hoi-Kwong Lo, Marcos Curty, and Bing Qi.
\newblock Measurement-device-independent quantum key distribution.
\newblock {\em Phys. Rev. Lett.}, 108:130503, March 2012.

\bibitem{Qi:TimeShift:2007}
Bing Qi, Chi-Hang~Fred Fung, Hoi-Kwong Lo, and Xiongfeng Ma.
\newblock Time-shift attack in practical quantum cryptosystems.
\newblock {\em Quant.~Inf.~Comput.}, 7:073, 2007.

\bibitem{Zhao:TimeshiftExp:2008}
Yi~Zhao, Chi-Hang~Fred Fung, Bing Qi, Christine Chen, and Hoi-Kwong Lo.
\newblock Experimental demonstration of time-shift attack against practical
  quantum key distribution systems.
\newblock {\em Phys. Rev. A}, 78:042333, 2008.

\bibitem{Fung:Remap:07}
Chi-Hang~Fred Fung, Bing Qi, Kiyoshi Tamaki, and Hoi-Kwong Lo.
\newblock Phase-remapping attack in practical quantum-key-distribution systems.
\newblock {\em Phys. Rev. A}, 75(3):032314, Mar 2007.

\bibitem{HKLO_PhaseRemap_NJP2010}
Feihu Xu, Bing Qi, and Hoi-Kwong Lo.
\newblock Experimental demonstration of phase-remapping attack in a practical
  quantum key distribution system.
\newblock {\em New Journal of Physics}, 12(11):113026, 2010.

\bibitem{Lydersen:Hacking:2010}
L.~Lydersen, C.~Wiechers, C.~Wittmann, D.~Elser, J.~Skaar, and V.~Makarov.
\newblock Hacking commercial quantum cryptography systems by tailored bright
  illumination.
\newblock {\em Nature photonics}, 4(10):686--689, 2010.

\bibitem{Wiechers:AftergateAttack:2011}
C~Wiechers, L~Lydersen, C~Wittmann, D~Elser, J~Skaar, Ch~Marquardt, V~Makarov,
  and G~Leuchs.
\newblock After-gate attack on a quantum cryptosystem.
\newblock {\em New Journal of Physics}, 13(1):013043, 2011.

\bibitem{Weier:DeadtimeAttack:2011}
Henning Weier, Harald Krauss, Markus Rau, Martin F\"urst, Sebastian Nauerth,
  and Harald Weinfurter.
\newblock Quantum eavesdropping without interception: an attack exploiting the
  dead time of single-photon detectors.
\newblock {\em New Journal of Physics}, 13(7):073024, 2011.

\bibitem{Jain:AttackExp:2011}
Nitin Jain, Christoffer Wittmann, Lars Lydersen, Carlos Wiechers, Dominique
  Elser, Christoph Marquardt, Vadim Makarov, and Gerd Leuchs.
\newblock Device calibration impacts security of quantum key distribution.
\newblock {\em Phys. Rev. Lett.}, 107:110501, Sep 2011.

\bibitem{Biham:ReverseEPR:1996}
E.~Biham, B.~Huttner, and T.~Mor.
\newblock Quantum cryptographic network based on quantum memories.
\newblock {\em Phys.~Rev.~A}, 54(4):2651, 1996.

\bibitem{MXF:MIQKD:2012}
Xiongfeng Ma and Mohsen Razavi.
\newblock Alternative schemes for measurement-device-independent quantum key
  distribution.
\newblock {\em Phys. Rev. A}, 86:062319, Dec. 2012.

\bibitem{MDIQKD_finite_PhysRevA2012}
Xiongfeng Ma, Chi-Hang~Fred Fung, and Mohsen Razavi.
\newblock Statistical fluctuation analysis for measurement-device-independent
  quantum key distribution.
\newblock {\em Phys. Rev. A}, 86:052305, Nov 2012.

\bibitem{Braunstein:MIQKD:2012}
Samuel~L. Braunstein and Stefano Pirandola.
\newblock Side-channel-free quantum key distribution.
\newblock {\em Phys. Rev. Lett.}, 108:130502, Mar 2012.

\bibitem{Razavi.Lutkenhaus.09}
Mohsen Razavi, Marco Piani, and Norbert L\"utkenhaus.
\newblock Quantum repeaters with imperfect memories: Cost and scalability.
\newblock {\em Phys. Rev. A}, 80:032301, Sept. 2009.

\bibitem{Razavi_SPIE}
M.~Razavi, K.~Thompson, H.~Farmanbar, Ma. Piani, and N.~L\"utkenhaus.
\newblock Physical and architectural considerations in quantum repeaters.
\newblock In {\em Proc. SPIE}, volume 7236, page 723603, San Jose, CA, 2009.

\bibitem{Razavi_IWCIT12}
Mohsen Razavi, Nicolo {Lo Piparo}, Christiana Panayi, and David~E. Bruschi.
\newblock Architectural considerations in hybrid quantum-classical networks
  (invited paper).
\newblock In {\em Iran Workshop on Communication and Information Theory
  (IWCIT)}, pages 1--7, Tehran, Iran, 2013.

\bibitem{MIT-NU}
S.~Lloyd, M.~S. Shahriar, J.~H. Shapiro, and P.~R. Hemmer.
\newblock Long distance, unconditional teleportation of atomic states via
  complete bell state measurements.
\newblock {\em Phys. Rev. Lett.}, 87:167903, Sep 2001.

\bibitem{Razavi.DLCZ.06}
Mohsen Razavi and Jeffrey~H. Shapiro.
\newblock Long-distance quantum communication with neutral atoms.
\newblock {\em Phys. Rev. A}, 73:042303, April 2006.

\bibitem{KBBBCDK_03}
A.~Kuzmich, W.~P. Bowen, A.~D. Boozer, A.~Boca, C.~W. Chou, L.-M. Duan, and
  H.~J. Kimble.
\newblock Generation of nonclassical photon pairs for scalable quantum
  communication with atomic ensembles.
\newblock {\em Nature}, 423:731, 2003.

\bibitem{Kuzmich_memory_05}
T.~Chaneli\`ere, D.~N. Matsukevich, S.~D. Jenkins, S.-Y. Lan, T.~A.~B. Kennedy,
  and A.~Kuzmich.
\newblock Storage and retrieval of single photons transmitted between remote
  quantum memories.
\newblock {\em Nature}, 438:833--836, 2005.

\bibitem{Zhao:Robust:2007}
Bo~Zhao, Zeng-Bing Chen, Yu-Ao Chen, J\"org Schmiedmayer, and Jian-Wei Pan.
\newblock Robust creation of entanglement between remote memory qubits.
\newblock {\em Phys. Rev. Lett.}, 98:240502, Jun 2007.

\bibitem{Pan:NatPhys:2012}
Xiao-Hui Bao, Andreas Reingruber, Peter Dietrich, Jun Rui, Alexander D\"uck,
  Thorsten Strassel, Li~Li, Nai-Le Liu, Bo~Zhao, and Jian-Wei Pan.
\newblock Efficient and long-lived quantum memory with cold atoms inside a ring
  cavity.
\newblock {\em Nat. Phys.}, 8:517--521, May 2012.

\bibitem{Rempe:Nature:2012}
Stephan Ritter, Christian N\"olleke, Carolin Hahn, Andreas Reiserer, Andreas
  Neuzner, Manuel Uphoff, Martin M\"ucke, Eden Figueroa, Joerg Bochmann, and
  Gerhard Rempe.
\newblock An elementary quantum network of single atoms in optical cavities.
\newblock {\em Nature}, 484:195--200, April 2012.

\bibitem{BB_84}
C.~H. Bennett and G.~Brassard.
\newblock Quantum cryptography: Public key distribution and coin tossing.
\newblock In {\em Proceedings of IEEE International Conference on Computers,
  Systems, and Signal Processing}, pages 175--179, Bangalore, India, 1984.
  IEEE, New York.

\bibitem{Kuklewicz:BrightEPR}
Christopher~E. Kuklewicz, Marco Fiorentino, Ga\'etan Messin, Franco N.~C. Wong,
  and Jeffrey~H. Shapiro.
\newblock High-flux source of polarization-entangled photons from a
  periodically poled {KTiOPO4} parametric down-converter.
\newblock {\em Phys. Rev. A}, 69:013807, Jan 2004.

\bibitem{MIT_polEntgSource}
Marco Fiorentino, Ga\'etan Messin, Christopher~E. Kuklewicz, Franco N.~C. Wong,
  and Jeffrey~H. Shapiro.
\newblock Generation of ultrabright tunable polarization entanglement without
  spatial, spectral, or temporal constraints.
\newblock {\em Phys. Rev. A}, 69:041801, Apr 2004.

\bibitem{Walmsley:PRL:2010}
K.~F. Reim, P.~Michelberger, K.~C. Lee, J.~Nunn, N.~K. Langford, and I.~A.
  Walmsley.
\newblock Single-photon-level quantum memory at room temperature.
\newblock {\em Phys. Rev. Lett.}, 107:053603, Jul 2011.

\bibitem{LoPiparo:2014}
Nicol\'o {Lo~Piparo} and Mohsen Razavi.
\newblock Measurement-device-independent quantum key distribution with
  imperfect sources and memories.
\newblock {\em in preparation}.

\bibitem{Lo:Decoy:2005}
Hoi-Kwong Lo, Xiongfeng Ma, and Kai Chen.
\newblock Decoy state quantum key distribution.
\newblock {\em Phys.~Rev.~Lett.~}, 94:230504, June 2005.

\bibitem{Lo:EffBB84:2005}
Hoi-Kwong Lo, H.~F. Chau, and M.~Ardehali.
\newblock Efficient quantum key distribution scheme and a proof of its
  unconditional security.
\newblock {\em Journal of Cryptology}, 18(2):133--165, 2005.

\bibitem{Chan:MIQKDexp:2013}
P.~Chan, J.~A. Slater, I.~Lucio-Martinez, A.~Rubenok, and W.~Tittel.
\newblock Modeling a measurement-device-independent quantum key distribution
  system.
\newblock {\em arXiv:1204.0738 [quant-ph]}, 2013.

\bibitem{Han:uncharSource_PRA2013}
Zhen-Qiang Yin, Chi-Hang~Fred Fung, Xiongfeng Ma, Chun-Mei Zhang, Hong-Wei Li,
  Wei Chen, Shuang Wang, Guang-Can Guo, and Zheng-Fu Han.
\newblock Measurement-device-independent quantum key distribution with
  uncharacterized qubit sources.
\newblock {\em Phys. Rev. A}, 88:062322, Dec. 2013.

\bibitem{Razavi_CohMeas_PRA09}
E.~Bocquillon, C.~Couteau, M.~Razavi, R.~Laflamme, and G.~Weihs.
\newblock Coherence measures for heralded single-photon sources.
\newblock {\em Phys. Rev. A}, 79:035801, March 2009.

\bibitem{Razavi_CohMeas_JPhysB09}
M~Razavi, I~S\"ollner, E~Bocquillon, C~Couteau, R~Laflamme, and G~Weihs.
\newblock Characterizing heralded single-photon sources with imperfect
  measurement devices.
\newblock {\em Journal of Physics B: Atomic, Molecular and Optical Physics},
  42(11):114013, 2009.

\bibitem{Telcordia_1550_1550}
N.~A. Peters, P.~Toliver, T.~E. Chapuran, R.~J. Runser, S.~R. McNown, C.~G.
  Peterson, D.~Rosenberg, N.~Dallmann, R.~J. Hughes, K.~P. McCabe, J.~E.
  Nordholt, and K.~T. Tyagi.
\newblock Dense wavelength multiplexing of 1550nm {QKD} with strong classical
  channels in reconfigurable networking environments.
\newblock {\em New J. Phys.}, 11:045012, April 2009.

\bibitem{Telcordia_1550_1310}
T.~E. Chapuran, P.~Toliver, N.~A. Peters, J.~Jackel, M.~S. Goodman, R.~J.
  Runser, S.~R. McNown, N.~Dallmann, R.~J. Hughes, K.~P. McCabe, J.~E.
  Nordholt, C.~G. Peterson, K.~T. Tyagi, L.~Mercer, and H.~Dardy.
\newblock Optical networking for quantum key distribution and quantum
  communications.
\newblock {\em New J. Phys.}, 11:105001, Oct. 2009.

\bibitem{Razavi_MulipleAccessQKD}
M.~Razavi.
\newblock Multiple-access quantum key distribution networks.
\newblock {\em IEEE Trans. Commun.}, 60(10):3071--3079, 2012.

\bibitem{BML_Squash_08}
Normand~J. Beaudry, Tobias Moroder, and Norbert L\"utkenhaus.
\newblock Squashing models for optical measurements in quantum communication.
\newblock {\em Phys. Rev. Lett.}, 101:093601, 2008.

\bibitem{Fung:2011:Squash}
Chi-Hang~Fred Fung, H.~F. Chau, and Hoi-Kwong Lo.
\newblock Universal squash model for optical communications using linear optics
  and threshold detectors.
\newblock {\em Phys. Rev. A}, 84:020303, Aug 2011.

\bibitem{STIRAP:PhysRevA.1989}
J.~R. Kuklinski, U.~Gaubatz, F.~T. Hioe, and K.~Bergmann.
\newblock Adiabatic population transfer in a three-level system driven by
  delayed laser pulses.
\newblock {\em Phys. Rev. A}, 40:6741--6744, Dec 1989.

\bibitem{Razavi.Memory.07}
Mohsen Razavi and Jeffrey~H. Shapiro.
\newblock Nonadiabatic approach to entanglement distribution over long
  distances.
\newblock {\em Phys. Rev. A}, 75:032318, 2007.

\bibitem{Multimode_Gisin_PRL07}
Christoph Simon, Hugues de~Riedmatten, Mikael Afzelius, Nicolas Sangouard, Hugo
  Zbinden, and Nicolas Gisin.
\newblock Quantum repeaters with photon pair sources and multimode memories.
\newblock {\em Phys. Rev. Lett.}, 98:190503, May 2007.

\bibitem{Gisin_Nature_AFC_2011}
Christoph Clausen, Imam Usmani, F\'elix Bussi\`eres, Nicolas Sangouard, Mikael
  Afzelius, Hugues de~Riedmatten, and Nicolas Gisin.
\newblock Quantum storage of photonic entanglement in a crystal.
\newblock {\em Nature}, 469:508--511, Jan. 2011.

\bibitem{Tittel_Nature_AFC_2011}
Erhan Saglamyurek, Neil Sinclair, Jeongwan Jin, Joshua~A. Slater, Daniel Oblak,
  F\'elix Bussi\`eres, Mathew George, Raimund Ricken, Wolfgang Sohler, and
  Wolfgang Tittel.
\newblock Broadband waveguide quantum memory for entangled photons.
\newblock {\em Nature}, 469:512--515, Jan. 2011.

\bibitem{Tittel:Ondemand_2013}
Neil Sinclair, Erhan Saglamyurek, Hassan Mallahzadeh, Joshua~A. Slater, Mathew
  George, Raimund Ricken, Morgan~P. Hedges, Daniel Oblak, Christoph Simon,
  Wolfgang Sohler, and Wolfgang Tittel.
\newblock A solid-state memory for multiplexed quantum states of light with
  read-out on demand.
\newblock {\em arXiv:1309.3202 [quant-ph]}, 2013.

\bibitem{ShorPreskill_00}
Peter~W. Shor and John Preskill.
\newblock Simple proof of security of the {BB84} quantum key distribution
  protocol.
\newblock {\em Phys.~Rev.~Lett.~}, 85(2):441, July 2000.

\bibitem{LoPiparo:2013}
Nicol\'o Lo~Piparo and Mohsen Razavi.
\newblock Long-distance quantum key distribution with imperfect devices.
\newblock {\em Phys. Rev. A}, 88:012332, Jul 2013.

\bibitem{Kuzmich_MultipleMem_PRL07}
O.~A. Collins, S.~D. Jenkins, A.~Kuzmich, and T.~A.~B. Kennedy.
\newblock Multiplexed memory-insensitive quantum repeaters.
\newblock {\em Phys. Rev. Lett.}, 98:060502, Feb 2007.

\bibitem{Razavi.Amirloo.10}
Jeyran Amirloo, Mohsen Razavi, and A.~Hamed Majedi.
\newblock Quantum key distribution over probabilistic quantum repeaters.
\newblock {\em Phys. Rev. A}, 82:032304, Sept. 2010.

\bibitem{Nam_NatPhot_93p_2013}
F.~Marsili, V.~B. Verma, J.~A. Stern, S.~Harrington, A.~E. Lita, T.~Gerrits,
  I.~Vayshenker, B.~Baek, M.~D. Shaw, R.~P. Mirin, and S.~W. Nam.
\newblock Detecting single infrared photons with 93\% system efficiency.
\newblock {\em Nat. Photon.}, 7:210--214, Feb. 2013.

\bibitem{Camacho_Natphot_2009}
Ryan~M. Camacho, Praveen~K. Vudyasetu, and John~C. Howell.
\newblock Four-wave-mixing stopped light in hot atomic rubidium vapour.
\newblock {\em Nat. Photon.}, 3:103--106, Jan. 2009.

\bibitem{Blatt:Natture:2012}
A.~Stute, B.~Casabone, P.~Schindler, T.~Monz, P.~O. Schmidt, B.~Brandst\"atter,
  T.~E. Northup, and R.~Blatt.
\newblock Tunable ion–photon entanglement in an optical cavity.
\newblock {\em Nature}, 485:482, May 2012.

\bibitem{Gisin:AFCmem_JLum2010}
A.~Amari, A.~Walther, M.~Sabooni, M.~Huang, S.~Kröll, M.~Afzelius, I.~Usmani,
  B.~Lauritzen, N.~Sangouard, H.~de~Riedmatten, and N.~Gisin.
\newblock Towards an efficient atomic frequency comb quantum memory.
\newblock {\em Journal of Luminescence}, 130(9):1579 -- 1585, 2010.

\bibitem{Gisin_AFC_polar_PRL_2012}
Christoph Clausen, F\'elix Bussi\`eres, Mikael Afzelius, and Nicolas Gisin.
\newblock Quantum storage of heralded polarization qubits in birefringent and
  anisotropically absorbing materials.
\newblock {\em Phys. Rev. Lett.}, 108:190503, May 2012.

\bibitem{Riedmatten_AFC_polar_PRL_2012}
Mustafa G\"undo\ifmmode~\breve{g}\else \u{g}\fi{}an, Patrick~M. Ledingham,
  Attaallah Almasi, Matteo Cristiani, and Hugues de~Riedmatten.
\newblock Quantum storage of a photonic polarization qubit in a solid.
\newblock {\em Phys. Rev. Lett.}, 108:190504, May 2012.

\bibitem{Guo_AFC_polar_PRL_2012}
Zong-Quan Zhou, Wei-Bin Lin, Ming Yang, Chuan-Feng Li, and Guang-Can Guo.
\newblock Realization of reliable solid-state quantum memory for photonic
  polarization qubit.
\newblock {\em Phys. Rev. Lett.}, 108:190505, May 2012.

\bibitem{PhysRevA.87.062335}
Sylvia Bratzik, Silvestre Abruzzo, Hermann Kampermann, and Dagmar Bru\ss{}.
\newblock Quantum repeaters and quantum key distribution: The impact of
  entanglement distillation on the secret key rate.
\newblock {\em Phys. Rev. A}, 87:062335, June 2013.

\end{thebibliography}


\end{document}